\documentclass[%
reprint,
amsmath,amssymb,
aps,
pra,
floatfix,
]{revtex4-2}

\usepackage[dvipdfmx]{graphicx}
\usepackage{dcolumn}
\usepackage{bm}

\usepackage{color}

\begin{document}
\setlength\textfloatsep{6pt}


\title{
Application and quantum properties of superpositions of oppositely squeezed states
}

\author{Hiroo Azuma$^{1}$}
\email{zuma@nii.ac.jp}
\author{William J. Munro$^{2}$}
\author{Kae Nemoto$^{2, 1}$}
\affiliation{$^1$Global Research Center for Quantum Information Science,
National Institute of Informatics, 2-1-2 Hitotsubashi, Chiyoda-ku, Tokyo 101-8430, Japan}
\affiliation{$^2$Okinawa Institute of Science and Technology Graduate University, Onna-son, Okinawa 904-0495, Japan}




\date{\today}

\begin{abstract}
We show that superpositions of oppositely squeezed states---non‑Gaussian Schr{\"{o}}dinger-cat-like states---exhibit enhanced nonclassical features and provide an entanglement advantage in the small‑squeezing regime.
These states possess photon‑number structures distinct from conventional coherent‑state cat states, and we analyze their Wigner functions and the entanglement generated when they are injected into a 50-50 beam splitter.
As a practical application, we demonstrate that they enable a high‑quality heralded single‑photon source whose second‑order intensity correlation function is smaller than that obtained from a pure two‑mode squeezed vacuum state.
We further propose a linear‑optical heralding scheme that approximates these superpositions without requiring strong Kerr nonlinearities.
Our results indicate that the superposition of oppositely squeezed states is a promising non‑Gaussian resource for quantum information processing, particularly for single‑photon generation.
\end{abstract}

\maketitle


\section{\label{section-introduction}Introduction}
Schr{\"{o}}dinger's famous ``cat'' thought experiment illustrated the counterintuitive nature of quantum superposition, particularly for macroscopic states
\cite{Schrodinger1935}.
In modern quantum optics, Schr{\"{o}}dinger cat states---superpositions of macroscopically distinct states---are not only theoretically intriguing
but also valuable for quantum information processing.
A well-studied example is the superposition of two coherent states with equal amplitude and opposite phase,
$|\alpha\rangle \pm |-\alpha\rangle$,
which are paradigmatic non-Gaussian states
\cite{Gerry2005,Ourjoumtsev2007,Lee2012}.
Such states can exhibit negative Wigner function values, a hallmark of nonclassicality, and can serve as qubit encodings for optical quantum computation
\cite{Gottesman2001,Ralph2003}.

While Gaussian states such as coherent and squeezed states are experimentally accessible and theoretically tractable,
non-Gaussian states are essential for surpassing the capabilities of Gaussian-only protocols
\cite{Duan2000,Simon2000,Bartlett2002}.
However, practical generation methods for non-Gaussian states remain limited.
For coherent-state cat states, deterministic schemes based on Kerr nonlinearities
\cite{Yurke1986}
are hindered by the weakness of available nonlinear media, while probabilistic photon-subtraction techniques
\cite{Dakna1997,Wegner2004,Ourjoumtsev2006,Neergaard-Nielsen2006,Wakui2007}
work effectively only for small amplitudes $\alpha$.

The applications of non-Gaussian states to quantum information processing have been extensively investigated by many researchers \cite{Walschaers2021}.
Several examples have been identified in which the use of non-Gaussian states yields significant differences from Gaussian-state-based approaches.
Reference \cite{Mika2022} experimentally demonstrated the generation of light with provably quantum non-Gaussian features from a warm atomic ensemble in a single-mode regime.
In Ref. \cite{Strinati2024}, the emergence of non-Gaussianity in the single quantum optical parametric oscillators with an applied external field was studied.
Along with the progress in application-oriented research on non-Gaussian states, studies have emerged that aim to generate non-Gaussian states not probabilistically but in a deterministic manner.
Reference \cite{Pasharavesh2024} showed a scheme for producing non-Gaussian states of few-photon light using deterministic photon subtraction.

This raises a natural question: can Schr{\"{o}}dinger cat states be formed not from coherent states, but from oppositely squeezed states
$|r;\pm\rangle\propto|r\rangle \pm |-r\rangle$,
where $|r\rangle$ represents a squeezed state with a real squeezing parameter $r$?
Such states are also non-Gaussian and possess photon-number structures distinct from coherent-state cats, potentially offering richer nonclassical properties.
They can, in principle, be generated via cross-Kerr interactions
\cite{Azuma2024},
but---as with coherent-state cats---their practical implementation is severely constrained by currently available nonlinearities.

In this work, we show that superpositions of oppositely squeezed states, which can be regarded as non‑Gaussian Schrödinger‑cat-like states, provide an entanglement advantage over the pure two‑mode squeezed vacuum state in the small‑squeezing regime.
We analyze their Wigner function and entanglement properties when injected into a beam splitter, clarifying the physical origin of this advantage.
As a potential application, we consider their use in a heralded single‑photon source.
We further propose an approximate generation method that avoids the impractical requirement of strong Kerr nonlinearities.
Taken together, these results indicate that the superpositions of oppositely squeezed states are promising non‑Gaussian resources for quantum information processing.

This paper is organized as follows.
Section~\ref{section-how-to-generate-superpositions-ideal}
provides an ideal scheme for generation of the superpositions of oppositely squeezed states.
In Sec.~\ref{section-application-single-photon-source}, we show a method for implementing a heralded single-photon source using the superpositions.
In Sec.~\ref{section-Wigner-function}, we numerically compute Wigner function.
Section~\ref{section-entanglement}
evaluates entanglement of two-mode states generated from $|r;\pm\rangle$.
Section~\ref{section-approximate-generation} provides an approximate scheme
for generating the superpositions
without Kerr nonlinearity.
Section~\ref{section-conclusion} provides conclusion.

\section{\label{section-how-to-generate-superpositions-ideal}Ideal generation of superpositions of oppositely
squeezed states}
A conceptually straightforward way to generate the superpositions
$|r;\pm\rangle$
is through a cross-Kerr interaction between a squeezed state and a coherent state, as proposed in
\cite{Azuma2024}.
The procedure is as follows.
We prepare an arbitrary state in mode 1,
$
|\psi\rangle_{1}
=
\sum_{n=0}^{\infty}c_{n}|n\rangle_{1}
$
where $\{|n\rangle: n=0, 1, ...\}$ are photon number states,
and a coherent state in mode 2,
$
|\alpha\rangle_{2}
$.
These are sent into a medium exhibiting a cross-Kerr nonlinearity, governed by the Hamiltonian
$
\hat{H}_{12}
=
\hbar\kappa
\hat{n}_{1}
\hat{n}_{2}
$
where $\hat{n}_{1}$ ($\hat{n}_{2}$) are the photonic number operators
for the $1$ ($2$) modes, respectively.
For the input
$
|\psi\rangle_{1}
|\alpha\rangle_{2}
$,
the evolution after time $\tau$ is
$
|\Psi_{\mbox{\scriptsize out}}(\tau)\rangle_{12}
=
\sum_{n=0}^{\infty}
c_{n}|n\rangle_{1}|\alpha e^{-i\kappa\tau n}\rangle_{2}
$.
If we choose
$
|\psi\rangle_{1}
$
to be a squeezed state
$
|\zeta\rangle_{1}
$
with
$
\zeta=r e^{i\phi}
$
and set
$
2\kappa\tau=\pi
$,
the output becomes
$
|\Psi_{\mbox{\scriptsize out}}\rangle_{12}
=
(1/2)
(|\zeta\rangle_{1}+|-\zeta\rangle_{1})
|\alpha\rangle_{2}
+
(1/2)
(|\zeta\rangle_{1}-|-\zeta\rangle_{1})
|-\alpha\rangle_{2}
$.
By homodyne detection distinguishing
$|\alpha\rangle_{2}$
and
$|-\alpha\rangle_{2}$,
we can herald the states
$|\zeta\rangle\pm|-\zeta\rangle$.
When $\zeta$ is real, that is, $\zeta=r$, including normalization factors, we can write the superpositions as
$|r;\pm\rangle={\cal N}_{\pm}^{-1/2}(r)(|r\rangle\pm|-r\rangle)$, where
$
{\cal N}_{\pm}(r)
=
2
[
1\pm
(\cosh r\sqrt{1+\tanh^{2}r})^{-1}
]
$.

In realistic materials,
$\kappa$ is extremely small.
For GaAs, using typical parameters
\cite{Boyd2020},
the required medium length to achieve
$
2\kappa\tau=\pi
$
exceeds 2 km, which is entirely impractical
especially when other parasitic effects are considered.
Although photonic-crystal slow-light techniques
\cite{Baba2008,Notomi2001,Inoue2005,Engelen2006,OFaolain2007}
can enhance the interaction, achieving a full $\pi$ phase shift at the single-photon level remains beyond current technology
\cite{Matsuda2009}.

\section{\label{section-application-single-photon-source}
Application of $|r;\pm\rangle$ to a heralded single-photon source}
In this section, we show an application of $|r;\pm\rangle$ to construction of the heralded single-photon source.
Its procedure is as follows.
First, we inject $|r;+\rangle$ and $|r;-\rangle$ into two input ports of a 50-50 beam splitter separately.
Second, detecting photons emitted from one output port of the beam splitter with a single-photon detector as a heralding signal, we obtain a single photon from the other output port of the beam splitter.

Now, we explain the first step.
We consider the beam splitter whose unitary operation is given by
$\hat{B}=\exp[(\pi/4)(\hat{a}^{\dagger}\hat{b}-\hat{a}\hat{b}^{\dagger})]$.
Injecting $|r;-\rangle_{a}$ and $|r;+\rangle_{b}$ into input ports of modes $a$ and $b$ of the beam splitter, respectively, we obtain
\begin{eqnarray}
&&
\hat{B}|r;-\rangle_{a}|r;+\rangle_{b} \nonumber \\
&=&
{\cal N}_{-}^{-1/2}(r)
{\cal N}_{+}^{-1/2}(r)
[\hat{S}_{a}(r)\hat{S}_{b}(r)
-
\hat{S}_{a}(-r)\hat{S}_{b}(-r) \nonumber \\
&&
+
\hat{S}_{ab}(-r)
-
\hat{S}_{ab}(r)]
|0\rangle_{a}|0\rangle_{b},
\end{eqnarray}
where we use formulae,
\begin{eqnarray}
\hat{B}\hat{S}_{a}(r)\hat{B}^{\dagger}
&=&
\hat{S}_{ab}(-r/2)\hat{S}_{a}(r/2)\hat{S}_{b}(r/2) \nonumber \\
&=&
\hat{S}_{a}(r/2)\hat{S}_{b}(r/2)\hat{S}_{ab}(-r/2), \nonumber \\
\hat{B}\hat{S}_{b}(r)\hat{B}^{\dagger}
&=&
\hat{S}_{ab}(r/2)\hat{S}_{a}(r/2)\hat{S}_{b}(r/2) \nonumber \\
&=&
\hat{S}_{a}(r/2)\hat{S}_{b}(r/2)\hat{S}_{ab}(r/2),
\label{formula-Sa-beam-splitter-0}
\end{eqnarray}
$\hat{S}_{a}(r)=\exp[(-r/2)(\hat{a}^{\dagger 2}-\hat{a}^{2})]$,
and
$\hat{S}_{ab}(r)=\exp[-r(\hat{a}^{\dagger}\hat{b}^{\dagger}-\hat{a}\hat{b})]$.
Here, we draw attention to the following facts.
The squeezed states $\hat{S}_{a}(r)|0\rangle_{a}$ and $\hat{S}_{a}(-r)|0\rangle_{a}$ are superpositions of the even photon-number states
$|0\rangle_{a}$, $|2\rangle_{a}$, $|4\rangle_{a}$, ....
By contrast, because
\begin{eqnarray}
&&
[\hat{S}_{ab}(-r)-\hat{S}_{ab}(r)]|0\rangle_{a}|0\rangle_{b} \nonumber \\
&=&
\frac{2}{\cosh r}\sum_{m=0}^{\infty}\tanh^{2m+1}r|2m+1\rangle_{a}|2m+1\rangle_{b},
\end{eqnarray}
$[\hat{S}_{ab}(-r)-\hat{S}_{ab}(r)]|0\rangle_{a}|0\rangle_{b}$ is a superposition of
$|1\rangle_{a}|1\rangle_{b}$, $|3\rangle_{a}|3\rangle_{b}$, $|5\rangle_{a}|5\rangle_{b}$, ....
Hence, if we perform the quantum nondemolition measurement of a parity of the number of the photons in mode $a$, we can distinguish
$[\hat{S}_{ab}(-r)-\hat{S}_{ab}(r)]|0\rangle_{a}|0\rangle_{b}$
from
$[\hat{S}_{a}(r)\hat{S}_{b}(r)
-
\hat{S}_{a}(-r)\hat{S}_{b}(-r)]|0\rangle_{a}|0\rangle_{b}$
\cite{Gerry2008},
and we obtain
\begin{equation}
|\Psi(r)\rangle_{ab}
=
\frac{\sqrt{\cosh(2r)}}{2\sinh r}
[\hat{S}_{ab}(-r)-\hat{S}_{ab}(r)]|0\rangle_{a}|0\rangle_{b}.
\end{equation}
If the state $|\Psi(r)\rangle_{ab}$ is realized, the probability that the numbers of photons of modes $a$ and $b$ are equal to $n_{a}$ and $n_{b}$, respectively, is given by
\begin{eqnarray}
&&
P(n_{a},n_{b}) \nonumber \\
&=&
\left\{
\begin{array}{ll}
0 & \mbox{for $n_{a}$ is even,}\\
\dfrac{\cosh(2r)}{\sinh^{2}r\cosh^{2}r}\tanh^{2n_{a}}r\delta_{n_{a},n_{b}} & \mbox{for $n_{a}$ is odd.}\\
\end{array}
\right.
\end{eqnarray}
Thus, if we detect $|1\rangle_{a}$, the single photon is emitted from port $b$ of the beam splitter with unit probability.

Next, we explain the second step.
If we can prepare the perfect single-photon detector, we can build an ideal heralded single-photon source with $|\Psi(r)\rangle_{ab}$.
However, from a practical viewpoint, we consider an imperfect ``click/no click" detector of efficiency $\eta$ by the positive operator-valued measure (POVM) $\{\hat{E},\hat{\bm{I}}-\hat{E}\}$, where
\begin{equation}
\hat{E}
=
\eta\sum_{k=1}^{\infty}(1-\eta)^{k-1}|k\rangle\langle k|,
\label{POVM-imperfect-detector-1}
\end{equation}
for a ``click" and $\hat{\bm{I}}-\hat{E}$ for ``no click".
If we place this detector on a path of mode $a$, the click probability is then
\begin{eqnarray}
&&
P_{\mbox{\scriptsize click}}(r,\eta) \nonumber \\
&=&
\eta\sum_{n_{a}=1}^{\infty}\sum_{n_{b}=1}^{\infty}
(1-\eta)^{n_{a}-1}P(n_{a},n_{b}) \nonumber \\
&=&
\eta\sum_{n=0}^{\infty}
(1-\eta)^{2n}\frac{\cosh(2r)}{\sinh^{2}r\cosh^{2}r}\tanh^{2(2n+1)}r.
\end{eqnarray}
The probability that the detector ``clicks" and the heralded single photon is actually emitted is given by
\begin{equation}
P_{\mbox{\scriptsize click},1}(r,\eta)
=
\eta
\frac{\cosh(2r)}{\sinh^{2}r\cosh^{2}r}\tanh^{2}r.
\end{equation}
By contrast, the corresponding probabilities for the pure two-mode squeezed vacuum state
$|\Psi_{\mbox{\scriptsize TMSV}}(r)\rangle_{ab}
=\sqrt{1-q^{2}}\sum_{n=0}^{\infty}q^{n}|n\rangle_{a}|n\rangle_{b}$
are given by
\begin{equation}
P_{\mbox{\scriptsize click,TMSV}}(r,\eta)
=
\eta(1-q^{2})\sum_{n=1}^{\infty}(1-\eta)^{n-1}q^{2n},
\end{equation}
\begin{equation}
P_{\mbox{\scriptsize click},1,\mbox{\scriptsize TMSV}}(r,\eta)
=
\eta(1-q^{2})q^{2},
\end{equation}
where $q=\tanh r$ and $r$ is a squeezing parameter.
Then, the generation rates of the heralded single photons are expressed by
$G(r,\eta)=P_{\mbox{\scriptsize click},1}(r,\eta)/P_{\mbox{\scriptsize click}}(r,\eta)$
and
$G_{\mbox{\scriptsize TMSV}}(r,\eta)=P_{\mbox{\scriptsize click},1,\mbox{\scriptsize TMSV}}(r,\eta)/P_{\mbox{\scriptsize click,TMSV}}(r,\eta)$
for $|\Psi(r)\rangle_{ab}$ and $|\Psi_{\mbox{\scriptsize TMSV}}(r)\rangle_{ab}$, respectively.

\begin{figure}[ht]
\begin{center}
\includegraphics[width=\linewidth]{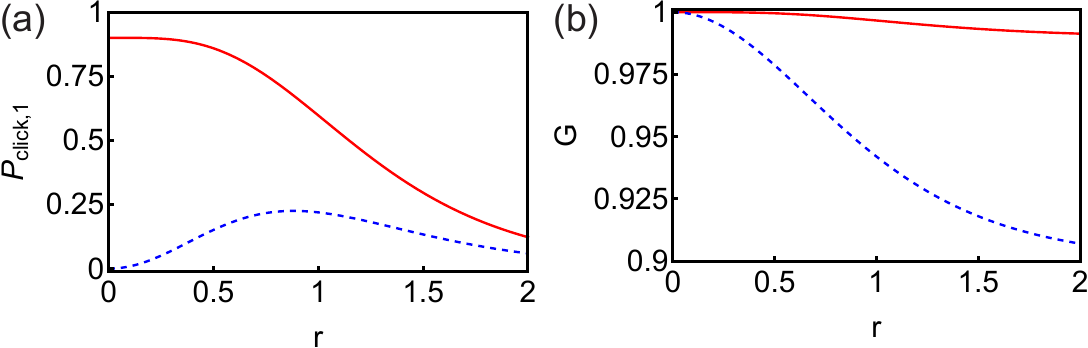}
\end{center}
\caption{
(a) Plots of $P_{\mbox{\scriptsize click},1}(r,\eta)$ and $P_{\mbox{\scriptsize click},1,\mbox{\scriptsize TMSV}}(r,\eta)$ as functions of $r$ with $\eta=0.9$.
The solid red and dashed blue curves represent $P_{\mbox{\scriptsize click},1}(r,\eta)$ and $P_{\mbox{\scriptsize click},1,\mbox{\scriptsize TMSV}}(r,\eta)$, respectively.
(b) Plots of generation rates of the single photons for $|\Psi(r)\rangle_{ab}$ and $|\Psi_{\mbox{\scriptsize TMSV}}(r)\rangle_{ab}$ with the imperfect single-photon detector as functions of $r$ with $\eta=0.9$.
The solid red and dashed blue curves represent the generation rates
$G(r,\eta)$ and $G_{\mbox{\scriptsize TMSV}}(r,\eta)$, respectively.
Both graphs (a) and (b) demonstrate that our proposed method is superior to the approach based on the pure two-mode squeezed vacuum state.
}
\label{figure01}
\end{figure}

In Fig.~\ref{figure01}(a), we plot $P_{\mbox{\scriptsize click},1}(r,\eta)$ and $P_{\mbox{\scriptsize click},1,\mbox{\scriptsize TMSV}}(r,\eta)$ as functions of $r$ with $\eta=0.9$.
Looking at Fig.~\ref{figure01}(a), we note that $P_{\mbox{\scriptsize click},1,\mbox{\scriptsize TMSV}}(r,\eta)$ is nearly equal to zero for $r=0$.
By contrast, $P_{\mbox{\scriptsize click},1}(r,\eta)$ is equal to $0.9$ for $r=0$.
In Fig.~\ref{figure01}(b), we plot generation rates of the single photons for $|\Psi(r)\rangle_{ab}$ and $|\Psi_{\mbox{\scriptsize TMSV}}(r)\rangle_{ab}$ with $\eta=0.9$, which are given by $G(r,\eta)$ and $G_{\mbox{\scriptsize TMSV}}(r,\eta)$, respectively.
Looking at Fig.~\ref{figure01}(b), we note that $|\Psi(r)\rangle_{ab}$ is preferable to $|\Psi_{\mbox{\scriptsize TMSV}}(r)\rangle_{ab}$ for implementing the heralded single-photon source in terms of the generation rate.

\begin{figure}[ht]
\begin{center}
\includegraphics[width=0.6\linewidth]{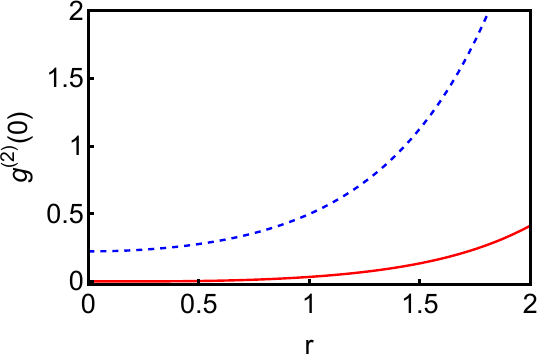}
\end{center}
\caption{
Plots of $g^{(2)}(0)$ for our proposed method and the pure two-mode squeezed vacuum state as functions of $r$ with $\eta=0.9$.
The solid red and dashed blue curves represent $g^{(2)}(0)$ for $|\Psi(r)\rangle_{ab}$ and $|\Psi_{\mbox{\scriptsize TMSV}}(r)\rangle_{ab}$, respectively.
In the method employing our proposed $|\Psi(r)\rangle_{ab}$, the value of $g^{(2)}(0)$ becomes nearly zero in the region $r\leq 0.5$,
despite the single-photon detector having the efficiency $0.9$.
This indicates that the proposed method realizes a heralded single-photon source with $g^{(2)}(0) \approx 0$ for $r \lesssim 0.5$ even with non-unit detector efficiency, and outperforms the source based on a pure two-mode squeezed vacuum state in this low-squeezing regime.}
\label{figure02}
\end{figure}

The second-order intensity correlation function $g^{(2)}(0)$ is often used as an indicator of the quality of a single-photon source.
Its definition is given by
\begin{equation}
g^{(2)}(0)
=
\frac{\langle n(n-1)\rangle}{\langle n\rangle^{2}}.
\end{equation}
For our proposed method, we obtain
\begin{eqnarray}
g^{(2)}(0)
&=&
-
\frac{2(1-\eta)^{2}\mbox{sech}(2r)\sinh^{4}r}{\eta[1+(1-\eta)^{2}\tanh^{4}r]^{2}} \nonumber \\
&&
\times
[-1+(1-\eta)^{2}\tanh^{4}r] \nonumber \\
&&
\times
[3+(1-\eta)^{2}\tanh^{4}r].
\label{g-2-0-our_proposed_method-0}
\end{eqnarray}
On the other hand, for the pure two-mode squeezed vacuum state,
\begin{equation}
g^{(2)}(0)
=
-3+\frac{2}{\eta}+\eta+(1-\eta)\cosh(2r).
\label{g-2-0-pure_two-mode_squeezed_vacuum-0}
\end{equation}
In Fig.~\ref{figure02}, we plot $g^{(2)}(0)$ given by Eqs.~(\ref{g-2-0-our_proposed_method-0}) and (\ref{g-2-0-pure_two-mode_squeezed_vacuum-0}) as functions of $r$ with $\eta=0.9$.
Looking at Fig.~\ref{figure02}, we note that $g^{(2)}(0)$ of our method is always smaller than that of the pure two-mode squeezed vacuum state for $0\le r\le 2$.
In particular, our proposed method realizes $g^{(2)}(0)\simeq 0$ for $r=0$ and $\eta=0.9$.
By contrast, the pure two-mode squeezed vacuum state reveals $g^{(2)}(0)=2/9$ for $r=0$ and $\eta=0.9$.
Thus, we can conclude that the heralded single-photon source based on our proposed method outperforms that produced by the pure two-mode squeezed vacuum state.

\section{\label{section-Wigner-function}Wigner functions of the superpositions of oppositely squeezed states}
\begin{figure*}[ht]
\begin{center}
\includegraphics[width=0.9\linewidth]{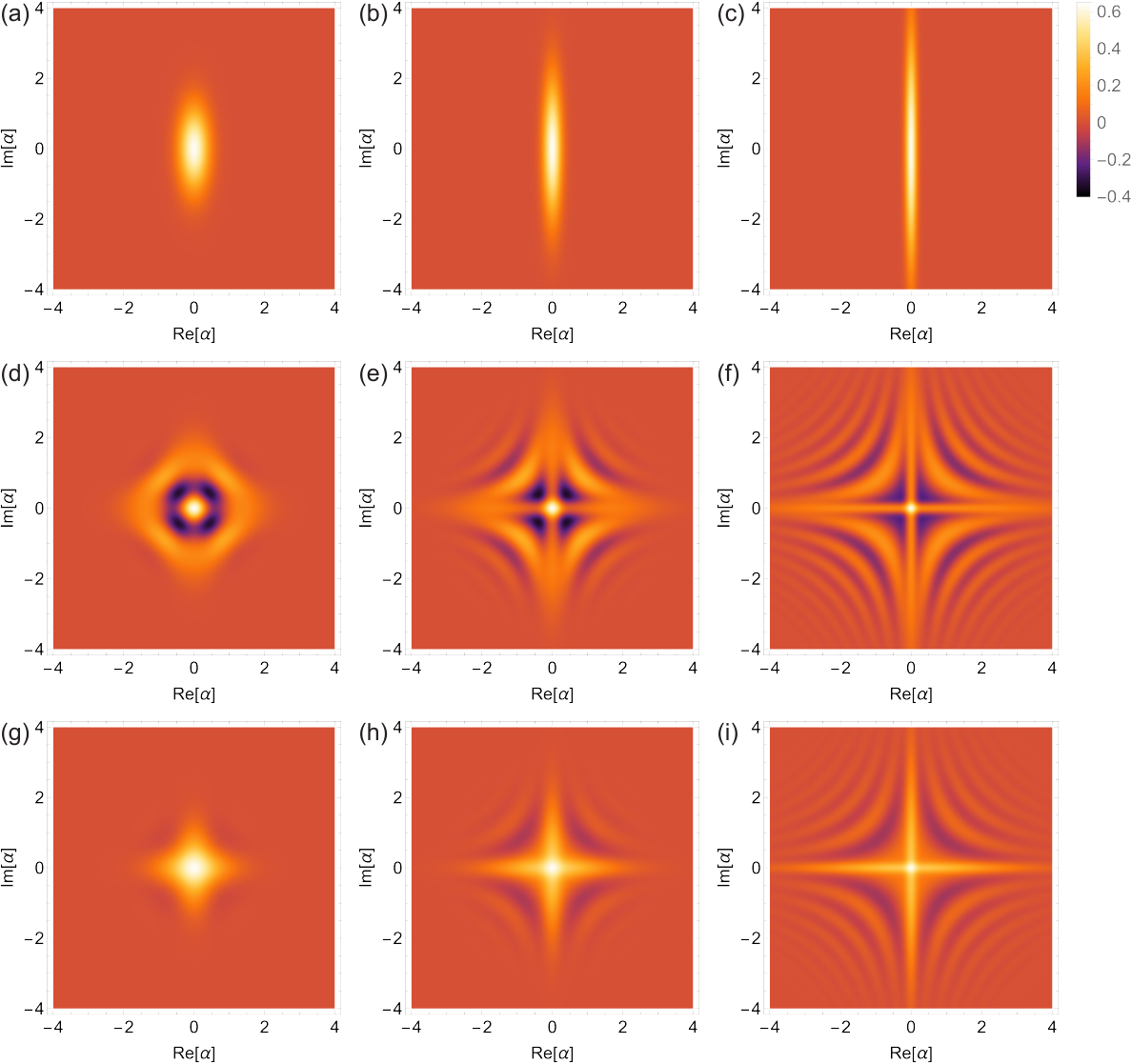}
\end{center}
\caption{Phase space plots of the Wigner function $W(\alpha)$ as a function of
$\mbox{Re}[\alpha]$ and $\mbox{Im}[\alpha]$.
(a) $|r\rangle$, $r=0.5$,
(b) $|r\rangle$, $r=1.0$,
(c) $|r\rangle$, $r=1.5$,
(d) $|r;-\rangle$, $r=0.5$,
(e) $|r;-\rangle$, $r=1.0$,
(f) $|r;-\rangle$, $r=1.5$,
(g) $|r;+\rangle$, $r=0.5$,
(h) $|r;+\rangle$, $r=1.0$,
(i) $|r;+\rangle$, $r=1.5$.
In graphs (a), (b), and (c), there is no region of $\alpha$ in which $W(\alpha)$ becomes negative.
In contrast, in graphs (d), (e), (f), (g), (h), and (i), there exist regions of $\alpha$ where $W(\alpha)$ takes negative values.
As seen in graphs (f) and (i), when $r$ is large, the value of $W(\alpha)$ oscillates between positive and negative with only slight changes in $\alpha$.
}
\label{figure03}
\end{figure*}

\begin{figure}[ht]
\begin{center}
\includegraphics[width=0.6\linewidth]{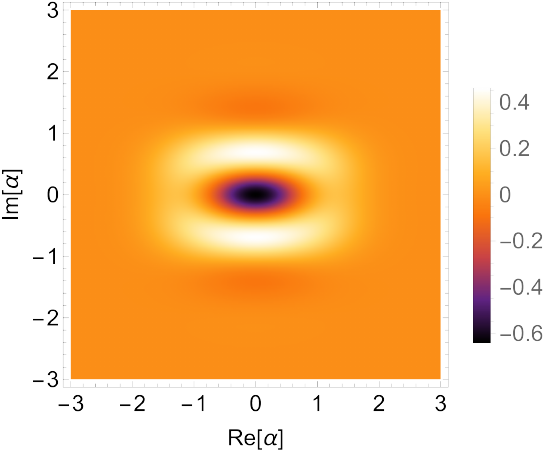}
\end{center}
\caption{Phase space plot of $W(\alpha)$ for the conventional coherent-state cat state
$\{2[1-\exp(-2a^{2})]\}^{-1/2}(|a\rangle-|-a\rangle)$ with $a=1$.
The graph has a negative value at the origin, $W(0)=-0.6366$.}
\label{figure04}
\end{figure}

Now let us move to the characterization of our superposition of oppositely squeezed states.
The Wigner function offers a complete phase-space representation of a quantum state and provides a direct signature of nonclassicality
through its negativity.
For a state $\rho$, the Wigner function is defined as
\begin{equation}
W(\alpha)
=
\frac{1}{\pi^{2}}
\int_{-\infty}^{\infty}
\int_{-\infty}^{\infty}
d^{2}\xi\:
\chi(\xi)
\exp(\alpha\xi^{*}-\alpha^{*}\xi),
\label{definition-Wigner-function-0}
\end{equation}
where
$\chi(\xi)
=
\mbox{Tr}
[\rho\hat{D}(\xi)]$
is the characteristic function
\cite{Barnett1997}.
For a single-mode squeezed vacuum $|r\rangle$ with real positive $r$, the Wigner function is a Gaussian ellipse in phase space,
$W(\alpha)
=
(2/\pi)
\exp[-2
(
\alpha_{\mbox{\scriptsize i}}^{2}e^{-2r}
+
\alpha_{\mbox{\scriptsize r}}^{2}e^{2r}
)
]$,
where
$\alpha_{\mbox{\scriptsize r}}=\mbox{Re}[\alpha]$
and
$\alpha_{\mbox{\scriptsize i}}=\mbox{Im}[\alpha]$
\cite{Walls1994}.
This function is always positive, reflecting the fact that squeezed vacua are Gaussian states.

When explicitly computing $W(\alpha)$ of $|r;\pm\rangle$, the following formula is more convenient than Eq.~(\ref{definition-Wigner-function-0}) \cite{Gerry2005}:
\begin{eqnarray}
W(\alpha)
&=&
\frac{2}{\pi}\sum_{n=0}^{\infty}(-1)^{n}
\langle r;\pm|\hat{D}(\alpha)|n\rangle\langle n|\hat{D}^{\dagger}(\alpha)|r;\pm\rangle \nonumber \\
&=&
\frac{2}{\pi}\frac{1}{{\cal N}_{\pm}(r)}
\sum_{n=0}^{\infty}(-1)^{n}
(\langle r|\pm\langle -r|)
\hat{D}(\alpha)|n\rangle \nonumber \\
&&
\times
\langle n|\hat{D}^{\dagger}(\alpha)
(|r\rangle\pm|-r\rangle).
\end{eqnarray}
Using a relation
\begin{eqnarray}
\langle n|\hat{D}(\alpha)|\xi\rangle
&=&
\frac{1}{\sqrt{\cosh r}}
\exp
\left[
-\frac{1}{2}|\alpha|^{2}-\frac{1}{2}\alpha^{*2}e^{i\theta}\tanh r
\right] \nonumber \\
&&
\times
\frac{[(1/2)e^{i\theta}\tanh r]^{n/2}}{\sqrt{n!}} \nonumber \\
&&
\times
H_{n}[\gamma(e^{i\theta}\sinh(2r))^{-1/2}],
\end{eqnarray}
for $\xi=e^{i\theta}r$ and $\gamma=\alpha\cosh r+\alpha^{*}e^{i\theta}\sinh r$ together with Eq.~(\ref{definition-Wigner-function-0})
where $H_{n}(x)$ is the Hermite polynomial of $x$,
we obtain the quantity $W(\alpha)$ of $|r;\pm\rangle$ as an infinite series.
By approximating the infinite series with sufficiently long finite ones, we can obtain the Wigner functions numerically.

For the superposition states $|r;\pm\rangle$,
we observe oscillatory interference fringes in phase space, which can drive the Wigner function negative.
This is a hallmark of non-Gaussianity and cannot occur for mixtures of Gaussian states.
Figure~\ref{figure03} illustrates this behavior.
As $r$ increases, the oscillations become finer in phase space, reflecting the increased separation between the two squeezed components.
Now, we compare $W(\alpha)$ of $|r;\pm\rangle$ with the Wigner function of the conventional coherent-state cat state.
The Wigner functions of the coherent state $|a\rangle$ and the cat state
$\{2[1-\exp(-2a^{2})]\}^{-1/2}(|a\rangle-|-a\rangle)$ with real $a$ are given by
$W(\alpha)=(2/\pi)\exp (-2|\alpha -a|^{2})$
and
\begin{eqnarray}
W(\alpha)
&=&
\frac{1}{\pi(1-e^{-2a^2})}
\left[
e^{-2|\alpha-a|^{2}}
+
e^{-2|\alpha+a|^{2}}
\right. \nonumber \\
&&
\left.
-2e^{-|\alpha|^{2}}\cos(4a\alpha_{\mbox{\scriptsize i}})
\right],
\end{eqnarray}
respectively.
We plot $W(\alpha)$ of the coherent-state cat state $[2(1-e^{-2})]^{-1/2}(|1\rangle-|-1\rangle)$ in Fig.~\ref{figure04}.
Exploring the plots, we note that the graph has a negative value at the origin, $W(0)=-0.6366$.
The Wigner functions of the coherent states $|a\rangle$ and $|-a\rangle$ have the maximum values at
$(\alpha_{\mbox{\scriptsize r}},\alpha_{\mbox{\scriptsize i}})=(a,0)$ and $(-a,0)$, respectively.
Thus, the superposition of $|a\rangle$ and $|-a\rangle$ with equal amplitude has interference at the origin,
a midpoint between $(a,0)$ and $(-a,0)$,
and its Wigner function becomes negative at that point.
By contrast, the Wigner functions of $|r;\pm\rangle$ have positive values at the origin as shown in Figs.~\ref{figure03}(d, e, f, g, h, i).
Because the Wigner functions of the squeezed states $|r\rangle$ and $|-r\rangle$ have positive values at the origin
and shapes of them are ellipses along $\alpha_{\mbox{\scriptsize r}}$- and $\alpha_{\mbox{\scriptsize i}}$-axes,
the superpositions $|r;\pm\rangle$ do not cause interference at the origin
and their Wigner functions become positive at that point.
The Wigner functions of $|r;\pm\rangle$ take negative values in the fringe around the origin, not at the origin.

The simplest index as a measure of the non-Gaussianity is the negativity volume of the Wigner function
\cite{Kenfack2004,Arkhipov2018}.
The definition of the negativity volume is given by
\begin{equation}
{\cal V}
=
\frac{1}{2}
\left(
\int
d^{2}\alpha
\,
|W(\alpha)|-1
\right).
\label{definition-negativity-volume-0}
\end{equation}
Thus, $|r;\pm\rangle$ have non-zero negativity volumes for $r>0$.
As seen from Eq.~(\ref{definition-negativity-volume-0}), evaluating the negativity volumes of $|r;\pm\rangle$ requires integrating the absolute value of $W(\alpha)$ over $\alpha$.
However, when performing the integration numerically,
it becomes difficult for the integral to converge to a stable value when the squeezing parameter $r$ is either very small or very large.
As a criterion for assessing whether the numerical integration results are reliable, we consider the following.
In general,
$\int
d^{2}\alpha
\,
W(\alpha)=1$
holds.
Therefore, we numerically evaluate $\int
d^{2}\alpha
\,
W(\alpha)$,
and if the resulting value lies between $0.975$ and $1.025$, we regard the numerical calculation as reliable.
As the specific numerical integration technique, we employ Romberg's method.

\begin{figure}[ht]
\begin{center}
\includegraphics[width=0.7\linewidth]{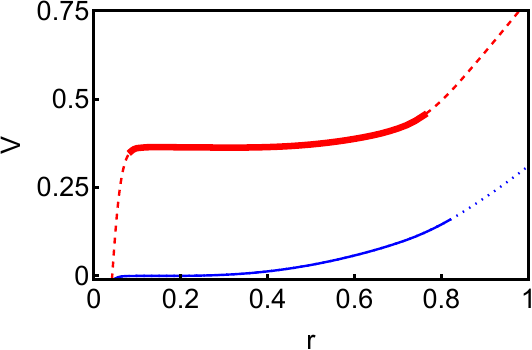}
\end{center}
\caption{
Plots of ${\cal V}$ of $|r;\pm\rangle$ as functions of the squeezing parameter $r$.
The thick solid red and thin solid blue curves show ${\cal V}$ of $|r;-\rangle$ for $0.086<r<0.76$ and
${\cal V}$ of $|r;+\rangle$ for $0.045<r<0.82$, respectively,
where the numerical results are reliable.
The thin dashed red and thin dotted blue curves represent extensions of ${\cal V}$
of $|r;-\rangle$ and $|r;+\rangle$ to regions in which the numerical calculations are not reliable.
As $r$ increases, ${\cal V}$ of $|r;\pm\rangle$ increases monotonically,
indicating that the non-Gaussianity becomes larger for larger $r$.
}
\label{figure05}
\end{figure}

From the above considerations, we find that the range of the squeezing parameter for which reliable numerical results can be obtained is
$0.086<r<0.76$ for $|r;-\rangle$ and $0.045<r<0.82$ for $|r;+\rangle$.
In Fig.~\ref{figure05}, we plot the negativity volume ${\cal V}$ of $|r;\pm\rangle$ as functions of $r$ over the above ranges.
Figure~\ref{figure05} shows that ${\cal V}$ of $|r;\pm\rangle$ monotonically increases as $r$ gets larger.
Thus, the non-Gaussianity can be considered to increase as the squeezing parameter $r$ becomes larger.
Moreover, ${\cal V}$ of $|r;-\rangle$ is always larger than ${\cal V}$ of $|r;+\rangle$.
Thus, the state $|r;-\rangle$ can be said to exhibit stronger non-Gaussianity than $|r;+\rangle$.

The negativity in $W(\alpha)$ for $|r;\pm\rangle$ confirms their strongly nonclassical nature.
This nonclassicality arises purely from superposition, not from squeezing alone:
even ordinary squeezed states can generate entanglement when combined at a beam splitter,
but they do not exhibit Wigner negativity.
The richer structure of $|r;\pm\rangle$ suggests potential advantages in tasks requiring high non-Gaussianity,
in particular, the heralded single-photon source analyzed in Sec.~\ref{section-application-single-photon-source}.

\section{\label{section-entanglement}Entanglement of two-mode states
generated from $|r;\pm\rangle$}
One way to assess the usefulness of $|r;\pm\rangle$ for continuous-variable quantum information is
to evaluate the entanglement they can produce when injected into a beam splitter.
Even Gaussian states, for example, the squeezed states, can generate entanglement in this way,
but the presence of strong non-Gaussianity can change both the amount and structure of the entanglement, potentially offering advantages for certain protocols.

As shown in Sec.~\ref{section-application-single-photon-source}, we can construct the state
$|\Psi(r)\rangle_{ab}\propto[\hat{S}_{ab}(-r)-\hat{S}_{ab}(r)]|0\rangle_{a}|0\rangle_{b}$
by injecting $|r;-\rangle_{a}$ and $|r;+\rangle_{b}$ into the 50-50 beam splitters.
Because the states considered here are pure bipartite states, we quantify entanglement using the entropy of the reduced density matrix.
We note that alternative entanglement measures, such as logarithmic negativity, may rank mixed or lossy states differently; however, the entropy provides a natural and unambiguous measure for the present comparison.
To quantify entanglement, we compute the entropy
\begin{eqnarray}
S(r)
&=&
-\mbox{Tr}[\rho_{a}(r)\log_{2}\rho_{a}(r)] \nonumber \\
&=&
-\sum_{m=0}^{\infty}p(r;m)\log_{2}p(r;m),
\label{definition-von-Neumann-entropy-0}
\end{eqnarray}
where
$\rho_{a}(r)
=
\mbox{Tr}_{b}
(|\Psi(r)\rangle_{ab}{}_{ab}\langle\Psi(r)|)$
and
\begin{equation}
p(r;m)
=
\frac{\cosh(2r)}{\sinh^{2}r\cosh^{2}r}\tanh^{2(2m+1)}r.
\end{equation}
For comparison, we also compute the entanglement of the following three states.
To obtain the first and second states,
we consider injecting $|r;\pm\rangle$ into mode $a$ of the 50-50 beam splitter with the other input mode $b$ in the vacuum state.
Using Eq.~(\ref{formula-Sa-beam-splitter-0}),
we obtain output states of the form
\begin{eqnarray}
|\Psi^{(\pm)}(r)\rangle_{ab}
&=&
{\cal N}_{\pm}(r)^{-1/2}
[
\hat{S}_{a}(r/2)
\hat{S}_{b}(r/2)
\hat{S}_{ab}(-r/2)|0\rangle \nonumber \\
&&
\pm
\hat{S}_{a}(-r/2)
\hat{S}_{b}(-r/2)
\hat{S}_{ab}(r/2)|0\rangle
],
\end{eqnarray}
and their entropies as $S^{(\pm)}(r)$.
As the third state, we consider the pure two-mode squeezed vacuum state
$|\Psi_{\mbox{\scriptsize TMSV}}(r)\rangle_{ab}$
given in Sec.~\ref{section-application-single-photon-source},
which can be generated experimentally by injecting two single-mode squeezed vacua into a beam splitter.
Its entropy is
\begin{equation}
S_{\mbox{\scriptsize TMSV}}(r)
=
-
\sum_{n=0}^{\infty}
(1-q^{2})
q^{2n}
\log_{2}
[(1-q^{2})q^{2n}],
\label{Sentropy-TMSV}
\end{equation}
where $q$ is associated with the squeezing parameter.

\begin{figure}[ht]
\begin{center}
\includegraphics[width=0.6\linewidth]{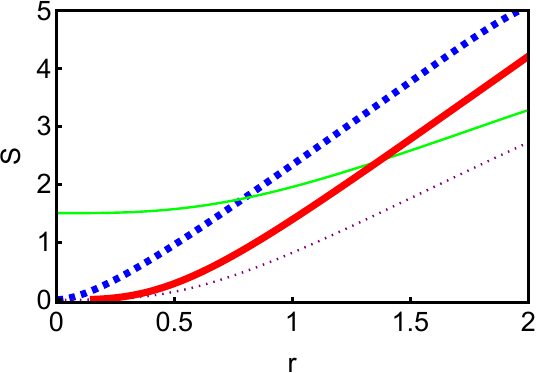}
\end{center}
\caption{
Plots of entropies as functions of the squeezing parameter $r$.
The thick solid red, thick dashed blue, thin solid green, and thin dotted purple curves represent $S(r)$, $S_{\mbox{\scriptsize TMSV}}(r)$,
$S^{(-)}(r)$, and $S^{(+)}(r)$, respectively.
For $0<r<0.787$, $S^{(-)}(r)>S_{\mbox{\scriptsize TMSV}}(r)$ holds.
For $0<r<1.37$, $S^{(-)}(r)>S(r)$ holds.
Thus, in the region of small $r$, $|r;-\rangle$ can serve as a good source of entanglement.
}
\label{figure06}
\end{figure}

Figure~\ref{figure06} plots the entropies as functions of $r$,
that is,
$S(r)$, $S_{\mbox{\scriptsize TMSV}}(r)$, $S^{(-)}(r)$, and $S^{(+)}(r)$.
Several features stand out:
For $r>0$, $S(r)$ is always smaller than $S_{\mbox{\scriptsize TMSV}}(r)$.
For small squeezing ($0<r<0.787$), $|\Psi^{(-)}(r)\rangle_{ab}$ produces more entanglement than the pure two-mode squeezed vacuum state.
Moreover, for $0<r<1.37$, $S^{(-)}(r)>S(r)$ holds.
This enhancement stems from the interference terms in $|r;-\rangle$, which increase the spread of photon-number correlations across modes.
For larger $r$, the advantage disappears, and the two-mode squeezed vacuum outperforms $|r;+\rangle$, $|r;-\rangle$, and $|\Psi(r)\rangle_{ab}$.

Here is an intuitive explanation why $S^{(-)}(r)>S_{\mbox{\scriptsize TMSV}}(r)$ holds in the small-$r$ regime.
From Eq.~(\ref{Sentropy-TMSV}), we obtain
\begin{equation}
S_{\mbox{\scriptsize TMSV}}(r)
=
\frac{(1-2\log r)r^{2}}{\log 2}+O(r^{3}).
\end{equation}
This implies $\lim_{r\to 0}S_{\mbox{\scriptsize TMSV}}(r)=0$.
On the other hand, $|r;-\rangle$ can be expanded in powers of $r$ as
\begin{equation}
|r;-\rangle
=
-
|2\rangle
-
\frac{1}{2}\sqrt{\frac{5}{2}}r^{2}|6\rangle+O(r^{3}).
\end{equation}
If we inject $|r;-\rangle_{a}$ and $|0\rangle_{b}$ into the 50-50 beam splitter $\hat{B}$,
the following state is emitted:
\begin{eqnarray}
\hat{B}|r;-\rangle_{a}|0\rangle_{b}
&=&
-
\frac{1}{2}|2\rangle_{a}|0\rangle_{b}
+
\frac{1}{\sqrt{2}}|1\rangle_{a}|1\rangle_{b} \nonumber \\
&&
-
\frac{1}{2}|0\rangle_{a}|2\rangle_{b}
+
O(r^{2}).
\end{eqnarray}
Tracing out the mode $b$ from the above state, we obtain the reduced density operator
\begin{equation}
\rho_{a}
=
\frac{1}{4}|0\rangle_{a}{}_{a}\langle 0|
+
\frac{1}{2}|1\rangle_{a}{}_{a}\langle 1|
+
\frac{1}{4}|2\rangle_{a}{}_{a}\langle 2|
+
O(r^{2}),
\end{equation}
whose entropy is $S^{(-)}(r)=3/2+O(r^{2})$.
Hence, when $r$ is close to zero, $|r;-\rangle$ generates larger entanglement than the two-mode squeezed vacuum state.

In the regime of a large squeezing parameter $r$,
the pure two-mode squeezed vacuum state exhibits a larger amount of entanglement.
The physical meaning of this is as follows.
In the limit of a large-but-finite squeezing regime of $r$, the pure two-mode squeezed vacuum state approaches the form
$\mbox{Const}.\times(|00\rangle_{ab}+|11\rangle_{ab}+|22\rangle_{ab}+...)$,
and this entangled state realizes ideal continuous-variable quantum teleportation.
In contrast, neither $|r;-\rangle$ nor $|\Psi(r)\rangle_{ab}$ achieves ideal quantum teleportation.
This difference is reflected in the value of the entanglement.

These results show that for low squeezing, the non-Gaussian superposition $|r;-\rangle$ can be a better entanglement resource
than a standard Gaussian two-mode squeezed state.
This could be relevant for protocols where experimental constraints limit achievable squeezing but allow for the preparation of modest-fidelity $|r;-\rangle$ states,
such as certain optical quantum communication.

\section{\label{section-approximate-generation}Approximate scheme for generating $|r;\pm\rangle$ without strong
Kerr nonlinearity}
Given the severe limitations of the ideal Kerr-based method, we now present a more practical approximate scheme for generating $|r;\pm\rangle$ that avoids relying on large optical
nonlinearities but is heralded in nature.
As explained in Sec.~\ref{section-application-single-photon-source}, the superpositions $|r;\pm\rangle$ are intended for implementing the heralded single-photon source, and therefore we focus on approximate states with small average photon numbers, up to six.
The basic idea is to construct $|r;+\rangle$ (and subsequently $|r;-\rangle$) from a two-mode squeezed vacuum state, ancillary vacuum modes,
beam splitters, small displacement operations, and photon counting as shown in Fig.~\ref{figure07} \cite{Lvovsky2002,Bimbard2010,Marek2011,Yukawa2013}.
The procedure is as follows:

\begin{figure}[ht]
\begin{center}
\includegraphics[width=0.7\linewidth]{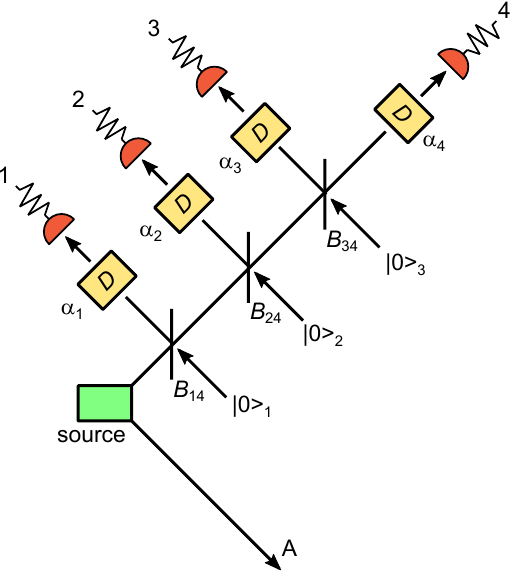}
\end{center}
\caption{Schematic illustration of implementation for generating $|r;+\rangle$
approximately.
The source emits the two-mode squeezed vacuum state for modes $4$ and A.
Next, preparing modes $1$, $2$, and $3$ as $|0\rangle_{1}|0\rangle_{2}|0\rangle_{3}$,
we apply the beam splitters
$\hat{B}_{14}$, $\hat{B}_{24}$, and $\hat{B}_{34}$
to modes $1$, $2$, $3$, and $4$.
Then, we apply the displacement operators
$\hat{D}_{j}(\alpha_{j})$ for $j=1, 2, 3$, and $4$
to the modes.
Finally, detecting single photons of modes $1$, $2$, $3$, and $4$,
we obtain $|r;+\rangle$ approximately for mode A.}
\label{figure07}
\end{figure}

\begin{enumerate}
\item
\textbf{Resource preparation:}
We begin with a two-mode squeezed vacuum state
whose squeezing parameter is given by $q$ in modes 4 and A.
Ancilla modes 1, 2, and 3 are prepared in the vacuum state.
\item
\textbf{Linear optical mixing:}
Three beam splitters couple mode 4 sequentially with modes 1, 2, and 3.
The beam splitter parameters are chosen to distribute photons coherently among these modes.
\item
\textbf{Small displacements:}
We then apply weak displacement operators $\hat{D}_{j}(\alpha_{j})$ to each mode $j=1,2,3$, and $4$, with $|\alpha_{j}| \ll 1$.
\item
\textbf{Photon detection and projection:}
Detecting exactly one photon in each of modes $1, 2, 3$, and $4$ projects mode A onto a superposition dominated by the
$|0\rangle$ and $|4\rangle$ Fock states.
By adjusting $\alpha_{j}$ and the squeezing parameter $q$ appropriately, this state approximates $|r;+\rangle$
with high fidelity.
\item
\textbf{Transforming to $|r;-\rangle$:}
Once $|r;+\rangle$ is prepared, a beam splitter transformation with an ancillary Fock state $|2\rangle$ can be used to engineer the relative phase
between even-photon components, yielding an approximate $|r;-\rangle$.
\end{enumerate}
Details of the above procedure are explained in Sec.~\ref{subsection-details-approximate-generation-scheme}.

This scheme uses only Gaussian resources (squeezed vacuum, displacements, beam splitters) plus photon counting, avoiding the need for strong nonlinearities.
The success probability that we obtain $|r;+\rangle$ approximately is proportional to $q^{8}$,
where $q$
is associated with the squeezing parameter of our initial two-mode squeezed vacuum state
$|\Psi_{\mbox{\scriptsize TMSV}}(s)\rangle_{4 \mbox{\scriptsize A}}=\sqrt{1-q^{2}}\sum_{n=0}^{\infty}q^{n}|n\rangle_{4}|n\rangle_{\mbox{\scriptsize A}}$,
so that $r$ and $q=\tanh s$ are independent parameters..

\begin{figure}[ht]
\begin{center}
\includegraphics[width=1.0\linewidth]{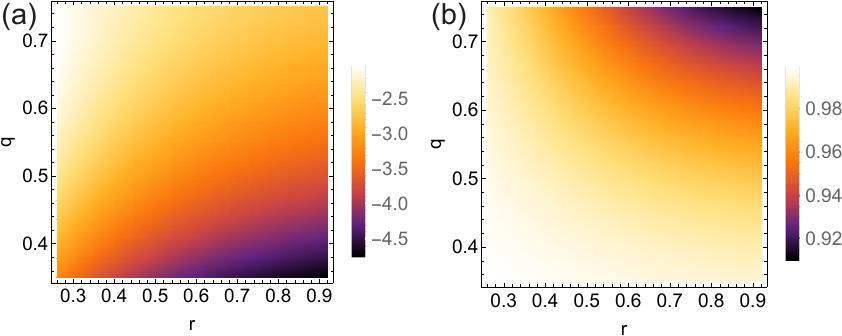}
\end{center}
\caption{(a)
Plot of $\log_{10}P$ where $P$ is the success probability that
all of the four unity-efficiency photon counters make click detection
with the experimental setup shown in Fig.~\ref{figure07}
as a function of $r$ and $q$.
(b)
Plot of the fidelity $F$ between the projected state of mode A in the experimental setup of Fig.~\ref{figure07}
and $|r;+\rangle$ as a function of $r$ and $q$.
In both (a) and (b), we set $0.26\leq r\leq 0.92$.
Looking at (a) and (b), we note that there is a tradeoff between $P$ and $F$ with respect to the parameter $q$.
Thus, it is necessary to choose an appropriate value of $q$.
}
\label{figure08}
\end{figure}

Here, we evaluate the probability $P$ that all of the four unity-efficiency photon counters detect single photons and the fidelity
$F=\langle r;+|\rho_{\mbox{\scriptsize A}}|r;+\rangle$
where $\rho_{\mbox{\scriptsize A}}$ represents a density matrix of mode A after photon counting of modes $1$, $2$, $3$, and $4$.
Figures~\ref{figure08}(a, b) show plots of $\log_{10}P$ and $F$ as functions of $r$ and $q$.
Figures~\ref{figure08}(a, b) show that $P$ increases and $F$ decreases as $q$ gets larger,
so that we recognize that there is a tradeoff between them.
To attain the balance between $P$ and $F$, we choose $q=0.5$.
Examining these plots,
we note that
$(2.09 \pm 0.04)\times 10^{-4} \leq P \leq (4.0 \pm 0.7)\times 10^{-3}$
and
$0.9767 \pm 4.0 \times 10^{-4} \leq F \leq 0.99667 \pm 2.0\times 10^{-5}$
hold with $q=0.5$.
Thus, if we emit the two-mode squeezed vacuum state with a generation rate of $1$ GHz
from the source in Fig.~\ref{figure07},
we obtain the approximate $|r;+\rangle$ at roughly a MHz rate.
These numbers are encouraging and suggest that the protocol may be feasible in practice.
However, this estimate is intended only as an order-of-magnitude illustration; the overall performance will ultimately be limited by setup-dependent losses and mode matching, which we do not model here.

Here, we have assumed that the generation rate of the two-mode squeezed vacuum state is $1$ GHz around.
We can realize this generation rate with cutting-edge technology.
Demonstrations of photon number resolving detectors with $1$ GHz count rate were reported in
\cite{Pearlman2005,Akiba2012,Pes2021,Wu2023,Shi2024}.
In particular, the authors of Ref.~\cite{Akiba2012} implemented a silicon photomultiplier detector that detected $2.6$ photons per optical pulse on average
with $1$ GHz count rate and $4.5${\%} detection efficiency at wave length $\lambda=775$ nm.
Thus, a generation rate $1$ GHz is feasible in experiments.

\begin{figure}[ht]
\begin{center}
\includegraphics[width=1.0\linewidth]{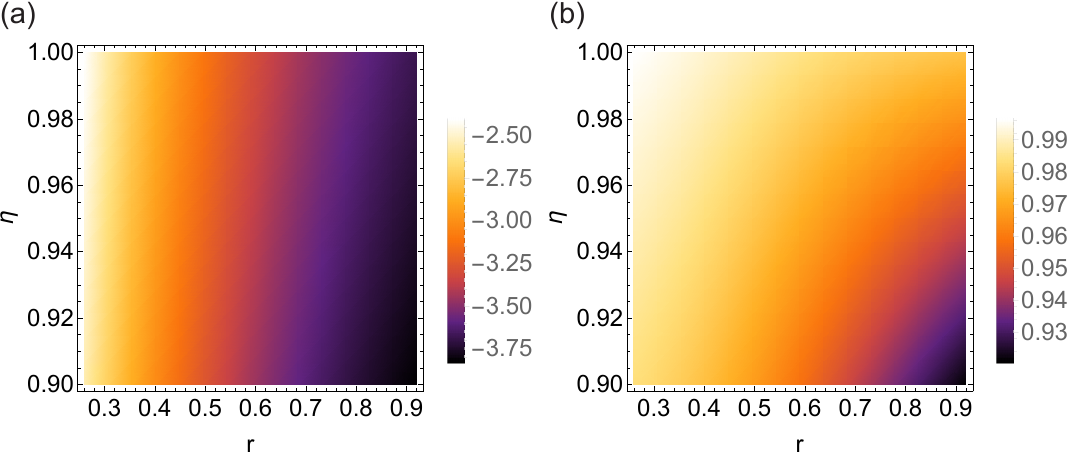}
\end{center}
\caption{
(a) Plot of $\log_{10}P$ where $P$ is the success probability that all of the four imperfect single-photon counters make click detection with the experimental setup shown in Fig.~\ref{figure07} as a function of the squeezing parameter $r$ and the efficiency of the single-photon counter $\eta$.
(b) Plot of the fidelity$F$ between the projected state of mode A in the experimental setup of Fig.~\ref{figure07} and $|r;+\rangle$
as a function of $r$ and $\eta$.
In both (a) and (b), we set $0.26\leq r\leq 0.92$ and $q=0.5$.
Graphs (a) and (b) show that $P$ and $F$ decrease monotonically as $\eta$ gets smaller for a fixed value $r$.
Thus, the efficiency of the single-photon detector is an important parameter in performing the experiment.
}
\label{figure09}
\end{figure}

Now, let us consider the case where the efficiency of the single-photon detectors is less than unity.
Assuming imperfect ``click/no click" detectors of efficiency $\eta$ as shown in Eq.~(\ref{POVM-imperfect-detector-1})
and setting $q=0.5$,
we evaluate $P$ and $F$.
Figures~\ref{figure09}(a, b) show plots of $\log_{10}P$ and $F$ as functions of $r$ and $\eta$.
Figures~\ref{figure09}(a, b) show that $P$ and $F$ decrease monotonically as $\eta$ gets smaller for a fixed value of $r$.
However, the reduction in $P$ is minor, indicating that it is not strongly influenced by variations in $\eta$.
In contrast, $F$ is strongly affected by the change in $\eta$.
For example, $F=0.9789$ for $(q,r,\eta)=(0.5,0.8,1)$,
and $F=0.9339$ for $(q,r,\eta)=(0.5,0.8,0.9)$.
Although the present parameter set yields high fidelities, the scheme may be somewhat sensitive to small deviations in these parameters.
A detailed robustness analysis is left for future work.

\subsection{\label{subsection-details-approximate-generation-scheme}Details of approximate generation of $|r;\pm\rangle$}
In this subsection, we explain details of how to generate $|r;\pm\rangle$ approximately
without nonlinear Kerr interaction.
Figure~\ref{figure07} gives the experimental setup for producing $|r;+\rangle$.
First of all, we prepare the two-mode squeezed vacuum state
$\hat{S}_{4\mbox{\scriptsize A}}(-s)
|0\rangle_{4}
|0\rangle_{\mbox{\scriptsize A}}$.
After operations of the step 3 in Sec.~\ref{section-approximate-generation},
we obtain the following state:
\begin{eqnarray}
&&
\hat{D}_{1}(\alpha_{1})\hat{D}_{2}(\alpha_{2})\hat{D}_{3}(\alpha_{3})\hat{D}_{4}(\alpha_{4})
\hat{B}_{34}\hat{B}_{24}\hat{B}_{14} \nonumber \\
&&
\times
\sqrt{1-q^{2}}
\sum_{n=0}^{\infty}q^{n}
|0\rangle_{1}|0\rangle_{2}|0\rangle_{3}|n\rangle_{4}|n\rangle_{\mbox{\scriptsize A}},
\label{two-mode-squeezed-vacuum-state-processed-0}
\end{eqnarray}
where $q=\tanh s$,
$s$ is the squeezing parameter, and it is an arbitrary real number.
The three beam splitters operate as follows:
\begin{eqnarray}
&&
\hat{B}_{14}:
\left(
\begin{array}{c}
\hat{a}_{1}' \\
\hat{a}_{4}'
\end{array}
\right)
=
\left(
\begin{array}{cc}
\sqrt{3}/2 & -1/2 \\
1/2 & \sqrt{3}/2
\end{array}
\right)
\left(
\begin{array}{c}
\hat{a}_{1} \\
\hat{a}_{4}
\end{array}
\right), \nonumber \\
&&
\hat{B}_{24}:
\left(
\begin{array}{c}
\hat{a}_{2}' \\
\hat{a}_{4}'
\end{array}
\right)
=
\left(
\begin{array}{cc}
\sqrt{2/3} & -1/\sqrt{3} \\
1/\sqrt{3} & \sqrt{2/3}
\end{array}
\right)
\left(
\begin{array}{c}
\hat{a}_{2} \\
\hat{a}_{4}
\end{array}
\right), \nonumber \\
&&
\hat{B}_{34}:
\left(
\begin{array}{c}
\hat{a}_{3}' \\
\hat{a}_{4}'
\end{array}
\right)
=
\left(
\begin{array}{cc}
1/\sqrt{2} & -1/\sqrt{2} \\
1/\sqrt{2} & 1/\sqrt{2}
\end{array}
\right)
\left(
\begin{array}{c}
\hat{a}_{3} \\
\hat{a}_{4}
\end{array}
\right).
\end{eqnarray}
The parameters of the displacement operators are small such that
$|\alpha_{j}|\ll 1$
for $j=1, 2, 3$, and $4$.
This means
$
\hat{D}_{j}(\alpha_{j})
\simeq
1+\alpha_{j}\hat{a}_{j}^{\dagger}-\alpha_{j}^{*}\hat{a}_{j}
$.
Finally, detecting single photons for modes $1$, $2$, $3$, and $4$,
we project the state of Eq.~(\ref{two-mode-squeezed-vacuum-state-processed-0})
onto $|1\rangle_{1}|1\rangle_{2}|1\rangle_{3}|1\rangle_{4}$.
Regarding the parameters $\{\alpha_{j}: j=1,2,3,4\}$ and $q$
as small amounts of the same magnitude,
we obtain the following terms up to fourth-order perturbation theory:
\begin{eqnarray}
&&
\sqrt{1-q^{2}}
\Bigg[
\alpha_{1}\alpha_{2}\alpha_{3}\alpha_{4}
|0\rangle_{\mbox{\scriptsize A}} \nonumber \\
&&
+
\frac{q}{2}
(\alpha_{1}\alpha_{2}\alpha_{3}
+
\alpha_{1}\alpha_{2}\alpha_{4}
+
\alpha_{1}\alpha_{3}\alpha_{4}
+
\alpha_{2}\alpha_{3}\alpha_{4})
|1\rangle_{\mbox{\scriptsize A}} \nonumber \\
&&
+
\frac{q^{2}}{2\sqrt{2}}
(\alpha_{1}\alpha_{2}
+
\alpha_{1}\alpha_{3}
+
\alpha_{1}\alpha_{4}
+
\alpha_{2}\alpha_{3}
+
\alpha_{2}\alpha_{4}
+
\alpha_{3}\alpha_{4})
|2\rangle_{\mbox{\scriptsize A}} \nonumber \\
&&
+
\frac{\sqrt{3}q^{3}}{4\sqrt{2}}
(\alpha_{1}
+
\alpha_{2}
+
\alpha_{3}
+
\alpha_{4})
|3\rangle_{\mbox{\scriptsize A}}
+
\frac{q^{4}}{4\sqrt{6}}
|4\rangle_{\mbox{\scriptsize A}}
\Bigg] \nonumber\\
&&
\times
|1\rangle_{1}
|1\rangle_{2}
|1\rangle_{3}
|1\rangle_{4}.
\label{projected-state-1111}
\end{eqnarray}
On the other hand, we can expand $|r;+\rangle$ as follows:
\begin{eqnarray}
|r;+\rangle
&\propto&
\frac{1}{2}(|r\rangle+|-r\rangle) \nonumber \\
&=&
\frac{1}{\sqrt{\cosh r}}
\left(
|0\rangle
+
\frac{\sqrt{3}}{2\sqrt{2}}\tanh^{2}r|4\rangle+...
\right).
\label{explicit-representation-r-plus-0}
\end{eqnarray}
Thus, we can obtain $|r;+\rangle$ approximately by adjusting
$\alpha_{j}$ ($j=1,2,3,4$) as follows:
\begin{eqnarray}
&&
\alpha_{1}\alpha_{2}\alpha_{3}\alpha_{4}
=
\frac{q^{4}}{6}\tanh^{-2}r, \nonumber \\
&&
\alpha_{1}\alpha_{2}\alpha_{3}
+
\alpha_{1}\alpha_{2}\alpha_{4}
+
\alpha_{1}\alpha_{3}\alpha_{4}
+
\alpha_{2}\alpha_{3}\alpha_{4}
=0, \nonumber \\
&&
\alpha_{1}\alpha_{2}
+
\alpha_{1}\alpha_{3}
+
\alpha_{1}\alpha_{4}
+
\alpha_{2}\alpha_{3}
+
\alpha_{2}\alpha_{4}
+
\alpha_{3}\alpha_{4}
=0, \nonumber \\
&&
\alpha_{1}
+
\alpha_{2}
+
\alpha_{3}
+
\alpha_{4}
=0.
\label{alpha-beta-gamma-delta-condition}
\end{eqnarray}
Here, we consider the following quartic equation:
\begin{equation}
(x-\alpha_{1})(x-\alpha_{2})(x-\alpha_{3})(x-\alpha_{4}) \nonumber \\
=
0.
\end{equation}
Because of Eq.~(\ref{alpha-beta-gamma-delta-condition}),
we obtain
\begin{equation}
x^{4}
+
\alpha_{1}\alpha_{2}\alpha_{3}\alpha_{4}
=
0,
\end{equation}
whose solutions are given by
\begin{equation}
x=c|q|e^{i\theta}
\quad
\mbox{for $\theta=\pi/4, 3\pi/4, 5\pi/4, 7\pi/4$,}
\end{equation}
with $c=6^{-1/4}\tanh^{-1/2}r$.
Thus, if we adjust the parameters as
\begin{eqnarray}
&&
\alpha_{1}=c|q|e^{i\pi/4},
\quad
\alpha_{2}=c|q|e^{i3\pi/4}, \nonumber \\
&&
\alpha_{3}=c|q|e^{i5\pi/4},
\quad
\alpha_{4}=c|q|e^{i7\pi/4},
\label{alphas-forms-1}
\end{eqnarray}
we can generate $|r;+\rangle$ approximately.
Because of Eqs.~(\ref{projected-state-1111}) and (\ref{alphas-forms-1}),
the amplitudes of the terms
$
|1\rangle_{1}
|1\rangle_{2}
|1\rangle_{3}
|1\rangle_{4}
|n\rangle_{\mbox{\scriptsize A}}
$
for $n=0$ and $4$ are of order $O(q^{4})$.
Hence, the probability that we obtain $|r;+\rangle$ is of order $O(q^{8})$.

Here, we draw attention to the following point.
Because the parameters
$\alpha_{j}$ ($j=1,2,3,4$)
and $q$ are in the same small order of magnitude,
the value $c$ is of order unity.
Thus, if we restrict $c$ to $3/4\leq c\leq 5/4$,
we obtain $0.2675\leq r\leq 0.9197$.
Under this condition, we can let the parameter $q$ be an arbitrarily small positive number.

Next, we explore how to obtain $|r;-\rangle$ approximately.
We have produced the following superposition already:
$
|\varphi\rangle
=
c_{0}|0\rangle
+
c_{4}|4\rangle
$.
Here, we consider a beam splitter $\hat{B}'$ whose transmittance is given by $T$.
Then, we apply that beam splitter to modes $a$ and $b$ whose state is given by
$|\varphi\rangle_{a}|2\rangle_{b}$ as
\begin{eqnarray}
|\varphi\rangle_{a}|2\rangle_{b}
&=&
\left(
c_{0}
+
\frac{c_{4}}{\sqrt{4!}}\hat{a}^{\dagger 4}
\right)
|0\rangle_{a}
\frac{1}{\sqrt{2!}}\hat{b}^{\dagger 2}
|0\rangle_{b} \nonumber \\
&\stackrel{\hat{B}'}{\to}&
\frac{c_{0}}{\sqrt{2!}}
(-\sqrt{1-T}\hat{a}^{\dagger}
+\sqrt{T}\hat{b}^{\dagger})^{2}
|0\rangle_{a}
|0\rangle_{b} \nonumber \\
&&
+
\frac{c_{4}}{\sqrt{4!}}
(\sqrt{T}\hat{a}^{\dagger}
+\sqrt{1-T}\hat{b}^{\dagger})^{4} \nonumber \\
&&
\times
(-\sqrt{1-T}\hat{a}^{\dagger}
+\sqrt{T}\hat{b}^{\dagger})^{2}
|0\rangle_{a}
|0\rangle_{b}.
\end{eqnarray}
Next, projecting that state onto $|0\rangle_{b}$ by observing no photons,
we obtain the following state probabilistically:
\begin{equation}
c_{0}
(1-T)|2\rangle_{a}
+
\sqrt{30}
c_{4}
T^{2}(1-T)|6\rangle_{a}.
\end{equation}
Hence, adjusting the value of $T$ appropriately,
we can obtain $|r\rangle-|-r\rangle$ approximately.

Now, we can evaluate the probability $P$ and fidelity $F$ introduced in Sec.~\ref{section-approximate-generation} with
the experimental setup shown in Fig.~\ref{figure07}.
We assume that the photon counters work
with the efficiency $\eta$, that is, the POVM of the counters is given by
$\{\hat{E},\hat{\bm{I}}-\hat{E}\}$ with Eq.~(\ref{POVM-imperfect-detector-1}).
We can numerically compute the probability that
all of the four photon counters make click detection by calculating
\begin{equation}
P
=
\langle \Psi|
(\hat{E}\otimes\hat{E}\otimes\hat{E}\otimes\hat{E}\otimes
\hat{\mbox{\boldmath $I$}})
|\Psi\rangle
\label{probability-0}
\end{equation}
where
$\hat{E}$ is given by Eq.~(\ref{POVM-imperfect-detector-1})
and
$|\Psi\rangle$ is given by Eq.~(\ref{two-mode-squeezed-vacuum-state-processed-0}).
Then, the fidelity $F$ is given by
\begin{equation}
F
=
\frac{1}{P}
\langle \Psi|
(\hat{E}\otimes\hat{E}\otimes\hat{E}\otimes\hat{E}\otimes
|r;+\rangle_{\mbox{\scriptsize A}}{}_{\mbox{\scriptsize A}}\langle r;+|)
|\Psi\rangle.
\end{equation}

For numerical calculations of $P$ and $F$ shown in Figs.~\ref{figure08}(a, b) and \ref{figure09}(a, b),
we assume that the initial two-mode squeezed vacuum state and the displacement operator are given by
sums of the first six terms as
$\sqrt{1-q^{2}}\sum_{n=0}^{5}q^{n}|n\rangle_{4}|n\rangle_{\mbox{\scriptsize A}}$
and
$\hat{D}(\alpha)
\simeq
\sum_{n=0}^{5}(1/n!)(\alpha\hat{a}^{\dagger}-\alpha^{*}\hat{a})^{n}$,
respectively.

\section{\label{section-conclusion}Conclusion}
We have analyzed the quantum properties of superpositions of oppositely squeezed states
$|r;\pm\rangle \propto |r\rangle \pm |-r\rangle$,
treating them as non-Gaussian Schr{\"{o}}dinger cat states.
As the practical application, we proposed the method for the implementation of the heralded single-photon source with $|r;\pm\rangle$ and the 50-50 beam splitter.
From numerical calculations of the second-order intensity correlation function, we found that our method is superior to the realization of the heralded single-photon source using the pure two-mode squeezed vacuum state.

Phase-space representations
revealed that both $|r;+\rangle$ and $|r;-\rangle$ are strongly nonclassical, exhibiting Wigner function negativity and fine-scale interference fringes.
Most notably, the entanglement analysis showed that when injected into the 50-50 beam splitter,
$|r;-\rangle$ produces more entanglement than a pure two-mode squeezed state for $0<r<0.787$.
This advantage vanishes for larger squeezing but suggests that $|r;-\rangle$ could be a superior resource in low-squeezing regimes
where Gaussian resources are limited.

From a practical standpoint, while the ideal cross-Kerr generation scheme is far beyond current nonlinear optical capabilities,
the linear-optical heralding method proposed in Sec.~\ref{section-approximate-generation} offers a potential route to approximate
$|r;\pm\rangle$ states without strong nonlinearities.
Future work should quantify the fidelity and success probability of this scheme under realistic losses.

In this paper, we have presented three results---Wigner-function negativity, a small-$r$ entanglement advantage, and the improved performance of a heralded single-photon source---which may appear, at first sight, to be unrelated.
However, these effects share a common physical origin in the interference structure of the superpositions $|r;\pm\rangle$.
In particular, the superposition modifies the photon-number distribution relative to that of a simple squeezed state, introducing interference terms that enhance photon-number correlations after the beam splitter.
This underlying structure simultaneously gives rise to the observed Wigner-function negativity and to the improved behavior in both the entanglement and heralded-photon protocols.

In summary, oppositely squeezed cat states combine the squeezing advantage of Gaussian states with the high non-Gaussianity of coherent-state cat states,
enabling high-quality single-photon production and stronger entanglement generation in certain regimes.
This dual nature makes them promising candidates for quantum information processing,
provided that experimentally feasible generation methods can be developed.

\section*{Acknowledgment}
This work was supported by MEXT Quantum Leap Flagship Program Grant No. JPMXS0120351339.

\end{document}